\documentclass[twoside]{article}
    \usepackage[a4paper]{geometry}
    
    \usepackage[maxbibnames=99, backend=biber, sorting=none]{biblatex}
    \usepackage[utf8]{inputenc}
    \usepackage[T1]{fontenc}
    \usepackage{RR}
    \usepackage[english]{babel}
    \usepackage[english]{isodate}
    \usepackage{indentfirst}
    \usepackage{hyperref}
    \usepackage{color, listings}
    \lstdefinelanguage{boogie}{
      keywords = {assert,assume,axiom,bool,break,bv0,bv1,call,complete,const,else,ensures,exists,false,finite,forall,free,function,goto,havoc,if,implementation,int,invariant,modifies,old,procedure,requires,return,returns,true,type,unique,var,where,while},
      sensitive=false,
      comment=[l]{//},
    }
    \definecolor{lightgray}{gray}{0.95}
    
    \lstdefinelanguage{json}{
        basicstyle=\footnotesize,
        breaklines=true,
        literate=
         *{:}{{{\color{blue}{:}}}}{1}
          {,}{{{\color{blue}{,}}}}{1}
          {\{}{{{\color{blue}{\{}}}}{1}
          {\}}{{{\color{blue}{\}}}}}{1}
          {[}{{{\color{blue}{[}}}}{1}
          {]}{{{\color{blue}{]}}}}{1},
    }

    \RRNo{9191}
    \RRdate{July 2018}
    \RRauthor{Sreeja Nair, Marc Shapiro}
    \authorhead{Sreeja Nair, Marc Shapiro}
    \RRtitle{}
    \RRetitle{Improving the ``Correct Eventual Consistency'' Tool}
    \RRresume{Dans les applications réparties s'exécutant dans un modèle de cohérence faible, le maintien des invariants est un problème difficile.
    L'outil CEC (\emph{Correct Eventual Consistency}) est destiné à aider le développeur d'application dans cette tâche.
    Il fournit des informations sur les erreurs lors des opérations simultanées et des suggestions sur comment et où synchroniser les opérations. 
    Dans ce rapport, nous présentons deux fonctionnalités de l'outil : la première fournit un contre-exemple lors du débogage et la seconde des suggestions permettant un meilleur contrôle de concurrence.
    }
    \RRabstract{
        Preserving invariants while designing distributed applications under weak consistency models is difficult.
      The CEC (\emph{Correct Eventual Consistency Tool}) is meant to aid the application designer in this task.
      It provides information about the errors during concurrent operations and suggestions on how and where to synchronize operations.
      This report presents two features of the tool: providing a counterexample for debugging and concurrency control suggestions. 
    }
    \RRkeyword{Consistency, Verification, Distributed applications}
    \RRmotcle{cohérence, vérification, applications distribuées}
    \RRprojets{DELYS}
    \RCParis
    
\begin{document}   
    \makeRR 
    \section{Introduction}

Reasoning about the correctness of distributed applications while ensuring high availability is a non-trivial task. 
CISE, introduced by \textcite{Gotsman:2016:CIS:2837614.2837625}, is a logic for reasoning about the correctness of a distributed application operating on top of a causally-consistent database.
In cases where the developer needs to control concurrent operations, he/she can use concurrency control named tokens.
According to the CISE logic, a distributed application is guaranteed to uphold its invariant if:
\begin{enumerate}
    \item Each operation is sequentially correct.
    \item The precondition of each operation is stable under any concurrent operation.
    \item Concurrent operations commute.
\end{enumerate}




Based on the CISE logic, a first tool was developed by \textcite{cisetool}.
This initial CISE tool is difficult because the user needs to use the low-level Z3 APIs directly \cite{mahsa}.
Subsequently, \textcite{goncalo} designed a second generation CISE tool, called Correct Eventual Consistency (CEC), which works on the same principle and provides a high-level developer friendly verification language, an extension of Boogie \cite{boogie} with specific annotations.
\par
An application written as a CEC specification consists of the following parts:
\begin{itemize}
  \item Data structures and properties
  \item Variables
  \item Invariants
  \item Operations with pre and post conditions
\end{itemize}

The tool first checks the specification for errors in the sequential specification.
The first step checks whether:
\begin{itemize}
  \item Each individual component of the specification is syntactically correct (\textit{syntax check}).
  \item Each operation satisfies the invariant individually (\textit{safety check}).
  \item The specification contains any contradictory clauses (\textit{anomaly check}).
  \item Every variable that is modified by an operation has a properly defined value when the operation terminates (\textit{completeness check}).
\end{itemize}

If the first phase passes, the analysis proceeds onto the second stage to check for concurrency. This stage verifies the CISE consistency conditions.
It checks every pair of operations to see whether:
\begin{itemize}
  \item The two operations of the pair have opposing preconditions including the tokens they acquire (\textit{opposition check}). 
  If so, they will never run concurrently. Otherwise (the two operations can run concurrently), the next two checks are applied.
  \item The precondition of one operation is preserved even under concurrent execution of the other operation with tokens (\textit{stability check}). 
  According to the CISE logic, this check ensures that the application's invariant is preserved. 
  \item The two operations commute (\textit{commutativity check}). This ensures convergence.
\end{itemize}



A first report on the evaluation of the tool is available \cite{nair:hal-01628719}. 
The current paper documents the improvements since the previous report in the following areas:
\begin{itemize}
    \item Counter example for failed verification.
    \item Optimized token generation.
\end{itemize}


    \section{Counterexample for failed verification}
\label{sec:counterex}

The previous report\cite{nair:hal-01628719} highlighted the lack of a counterexample when a verification fails.
The tool informed the user about the type of check that failed, it was difficult for the user to understand why.
The improved version of the tool now provides a counterexample
with the following information:
\begin{itemize}
    \item The statement that failed.
    \item The values of the parameters of the operation.
    \item The values of all the variables in the failed expression.
\end{itemize}

Let us illustrate with an example of a bank account. It has two operations - deposit and withdraw.
Suppose that the developer has specified a precondition for withdraw operation, but not for deposit.
\begin{lstlisting}[language=boogie]
@init
type Client;

@variables
var balances : [Client]int;

@equals [Client]int @as forall c: Client :: @this[c] == @other[c];

@invariant
forall c: Client :: balances[c] >= 0;

@operations

procedure deposit(accountId: Client, amount: int)
    modifies balances;
    ensures
        forall c: Client ::
        ((c != accountId) ==> (balances[c] == old(balances)[c])) &&
        ((c == accountId) ==> (balances[accountId] == old(balances)[accountId] + amount));

procedure withdraw(accountId: Client, amount: int)
    modifies balances;
    requires balances[accountId] - amount >= 0;
    ensures
        forall c: Client ::
        ((c != accountId) ==> (balances[c] == old(balances)[c])) &&
        ((c == accountId) ==> (balances[accountId] == old(balances)[accountId] - amount));

\end{lstlisting}

The \textcolor{blue}{\texttt{@init}} includes the initialization section of the specification. This sections contains datatype declarations, axioms and function declarations. In our case, there is a single datatype declaration here. The next section, \textcolor{blue}{\texttt{@variable}} contains all the variables used in the specification. The variable we use here is a mapping from each \texttt{Client} to their balance (which is an integer). The section \textcolor{blue}{\texttt{@equals}} gives the guidance to the tool about how to compare two variables for equality. In this specification, two variables of type \texttt{[Client]int} as considered equal if each \texttt{Client} has the same balance. The application invariant here is that the balances should always be non-negative. \textcolor{blue}{\texttt{@invariant}} states this application invariant and \textcolor{blue}{\texttt{@operations}} contains a list of operations of the application. Each operation will have a precondition which should be true at the origin replica (denoted by requires clause) and the effector function which is the result of executing the operation (denoted by the ensures clause).

The tool performs an analysis on this specification. In this case the safety test fails during the first stage (verifying the sequential specification).
The tool returns information on the assertion that failed, and the values of the variables involved in the assertion.
It also provides the values of the parameters of the failing operation.
Below is the corresponding output of the failure.
\begin{lstlisting}
############## SPECIFICATION CORRECTION TESTS ##############

--------------BASE VERIFICATION--------------
IMPORTS TEST: (PASSED)
INITIALIZER TEST: (PASSED)
VARIABLES TEST: (PASSED)
INVARIANTS TEST: (PASSED)
deposit OPERATION TEST: (PASSED)
withdraw OPERATION TEST: (PASSED)
FULL TEST: (PASSED)

--------------SEQUENTIAL VERIFICATION--------------
deposit SAFETY TEST: (FAILED)
DEBUG INFO :
The statements which failed are :::
ensures (forall c: Client :: balances[c] >= 0);
assume(forall c: Client :: balances[c] >= 0);
The value of accountId is T@Client!val!0.
The value of balances is |T@[Client]Int!val!1|.
The value of amount is (- 1200).
The value of c is .
The value of Client is .

withdraw SAFETY TEST: (PASSED)
deposit ABSURD TEST: (PASSED)
withdraw ABSURD TEST: (PASSED)
deposit COMPLETENESS TEST: (PASSED)
withdraw COMPLETENESS TEST: (PASSED)
FAILED LOGIC SPECIFICATION CORRECTION TEST
EXECUTION STOPPED
\end{lstlisting}

The first section of the result, the base verification, performs a syntax check for each part of the specification. The second part, the sequential verification checks whether the specification is safe sequentially.

The result tells that the tool was unable to verify that deposit operation ensures a positive balance, the application invariant.
The parameters of the failing deposit operation are \texttt{amount} and \texttt{accountId}. 
The value of \texttt{amount} is -1200, a negative value.
The developer takes this as a hint that negative values for \texttt{amount} are problematic. To test this hypothesis, he adds a precondition to indicate that \texttt{amount} should be positive. Then the specification looks like the following:
\begin{lstlisting}[language=boogie]
@init
type Client;

@variables
var balances : [Client]int;

@equals [Client]int @as forall c: Client :: @this[c] == @other[c];

@invariant
forall c: Client :: balances[c] >= 0;

@operations

procedure deposit(accountId: Client, amount: int)
    modifies balances;
    requires amount >= 0;
    ensures
        forall c: Client ::
        ((c != accountId) ==> (balances[c] == old(balances)[c])) &&
        ((c == accountId) ==> (balances[accountId] == old(balances)[accountId] + amount));

procedure withdraw(accountId: Client, amount: int)
    modifies balances;
    requires balances[accountId] - amount >= 0;
    ensures
        forall c: Client ::
        ((c != accountId) ==> (balances[c] == old(balances)[c])) &&
        ((c == accountId) ==> (balances[accountId] == old(balances)[accountId] - amount));

\end{lstlisting}

Now we rerun the tool with the corrected specification. 
We can see that the sequential safety test passes and we get the following result.
\begin{lstlisting}
############## SPECIFICATION CORRECTION TESTS ##############

--------------BASE VERIFICATION--------------
IMPORTS TEST: (PASSED)
INITIALIZER TEST: (PASSED)
VARIABLES TEST: (PASSED)
INVARIANTS TEST: (PASSED)
deposit OPERATION TEST: (PASSED)
withdraw OPERATION TEST: (PASSED)
FULL TEST: (PASSED)

--------------SEQUENTIAL VERIFICATION--------------
deposit SAFETY TEST: (PASSED)
withdraw SAFETY TEST: (PASSED)
deposit ABSURD TEST: (PASSED)
withdraw ABSURD TEST: (PASSED)
deposit COMPLETENESS TEST: (PASSED)
withdraw COMPLETENESS TEST: (PASSED)
############## CONSISTENCY TESTS ##############

--------------PAIR OPPOSITION VERIFICATION--------------
deposit deposit OPPOSITION TEST: (PASSED)
deposit withdraw OPPOSITION TEST: (PASSED)
withdraw withdraw OPPOSITION TEST: (PASSED)

--------------PAIR STABILITY VERIFICATION--------------
deposit deposit STABILITY TEST: (PASSED)
deposit withdraw STABILITY TEST: (PASSED)
withdraw withdraw STABILITY TEST: (FAILED)
DEBUG INFO :
The statements which failed are :::
requires (balances[accountId] - amount >= 0);
assume(forall c: Client :: balances[c] >= 0);
The value of accountId is T@Client!val!0,T@Client!val!0.
The value of balances is |T@[Client]Int!val!0|.
The value of amount is 797,797.
The value of c is .
The value of Client is .


--------------PAIR COMMUTATIVITY VERIFICATION--------------
deposit deposit COMMUTATIVITY TEST: (PASSED)
deposit withdraw COMMUTATIVITY TEST: (PASSED)
\end{lstlisting}

The stability test failed for concurrent \texttt{withdraw} operations.
Indeed, the precondition of the second \texttt{withdraw} is not stable under the execution of the first.
We can see that he values of \texttt{accountId} are the same for both operations.
This a hint for the developer to insert concurrency control for concurrent \texttt{withdraw} operations for the same \texttt{accountId}.

    \section{Synthesis of concurrency control}
\label{sec:tokengen}
The tool now leverages the counterexample obtained in Section \ref{sec:counterex} to suggest a concurrency control token. 

Let us continue with our bank account example. 
From the counter example, the tool infers two possible restrictions:
\begin{itemize}
    \item The \texttt{amount}s should be different
    \par
    and/or
    \item The \texttt{accountId}s should be different
\end{itemize}

The tool automatically reruns the analysis with these added restrictions. 
Even when the two \texttt{amount}s are different, the analysis still violates the stability check. 
Therefore this is not a fruitful excercise.
When the \texttt{accountId}s are constarined to be different the specification passes the stability test. 
This shows that concurrent withdraws to different accounts are ok, and suggests that concurrency should be disallowed for the same \texttt{accountId}.
The CISE abstract for concurrency control is called a token.

The output of the token generator is :
\begin{lstlisting}
--------------COUNTER EXAMPLE GUIDED SOLVER--------------
withdraw(accountId: Client, amount: int) withdraw(accountId: Client, amount: int) COUNTER EXAMPLE GUIDED SOLVER: (PASSED)
Solutions:
( @1 != @3 )

--------------TOKEN MODEL--------------
Tokens per operations:
withdraw: withdraw_accountId

Token conflicts:
withdraw_accountId: withdraw_accountId
\end{lstlisting}

The first part of the output tells that in the list of parameters, the first parameter should not be equal to the third one.
The token model presents the same information grouped in the form of tokens needed for each operation and conflicts for each token. 
This complies with the token specifications in CISE.
The token model tells that \texttt{withdraw} operation needs to acquire a token for each \texttt{account\_id}.
For a developer, this means the \texttt{withdraw} operations operating on one \texttt{account\_id} need to synchronize.





    \section{Conclusion}

This report summarises some recent improvements made on the Correct Eventual Consistency (CEC) tool. 
The tool now provides more comprehensible counter examples.
The tool also suggests tokens by utilising the information from the counter examples in an optimized fashion.

Currently the tool is providing suggestions only based on imposing inequality restrictions on the parameters. 
This can be improved considering more relations between parameters.
The next step would be to develop a complementary tool for analysing the applications which use state-based update propagation mechanism.

    \section*{Acknowledgement}
    The authors would like to thank Carla Ferreira, Gustavo Petri, Gon\c{c}alo Marcelino and Yann Thierry-Mieg for their inputs.
This research is supported in part 
       by the RainbowFS project
       (\emph{Agence Nationale de la Recherche}, France, number 
       ANR-16-CE25-0013-01) %
   and by European H2020 project
    \href{https://www.lightkone.eu/}{732\,505 LightKone} (2017--2020).%

\printbibliography

\end{document}